\documentclass[%
reprint,
superscriptaddress,
amsmath,amssymb,
aps,
prl,
floatfix,
twocolumn
]{revtex4-2}

\usepackage{comment}
\usepackage{graphicx}
\usepackage{dcolumn}
\usepackage{bm}
\usepackage [latin1]{inputenc}

\begin{document}


 \title{Fluctuation Theorem as a special case of Girsanov Theorem}
	
	\author{Annwesha Dutta}%
	\affiliation{%
School of Computational and Integrative Sciences, Jawaharlal Nehru University, New Delhi, India\\
	}%
	
	\author{Saikat Sarkar}%
	\email{saikat@iitd.ac.in}
	\affiliation{%
Department of Civil Engineering, Indian Institute of Technology (IIT) Delhi, New Delhi, India\\
	}%

	\date{\today}
	
	\begin{abstract}
Stochastic thermodynamics is an important development in the direction of finding general thermodynamic principles for non-equilibrium systems. We believe stochastic thermodynamics has the potential to benefit from the measure-theoretic framework of stochastic differential equations. Towards this, in this work, we show that Fluctuation Theorem (FT) is a special case of the Girsanov theorem, which is an important result in the theory of stochastic differential equations. We report that by employing Girsanov transformation of measures between the forward and the reversed dynamics of a general class of Langevin dynamic
systems,  we arrive at the Integral Fluctuation Relation. Following the same approach, we derive the FT also for the overdamped case. Our derivation is applicable to both transient and steady state conditions and can also incorporate diffusion coefficients varying as a function of state and time. We expect that the proposed method will be an easy route towards deriving the FT irrespective of the complexity and non-linearity of the system. 

	\end{abstract}
	
	\maketitle
\section{Introduction}
One of the primary objectives of statistical physics is to develop a robust framework for nonequilibrium systems as an extension of its equilibrium counterpart.
Stochastic thermodynamics is an important development in this direction which systematically defines thermodynamic quantities for individual fluctuating trajectories of non-equilibrium mesoscopic systems that are coupled to bath, ranging from biological \cite{Monge_2018} to quantum systems \cite{Talkner2009,HernndezGmez2021,Ciliberto2017-fu}. 
 Over the last couple of decades, researchers have derived several crucial results in stochastic thermodynamics, among which fluctuation theorem (FT) is a central result \cite{seifert2012stochastic}. It provides us with an equality closure for the physical processes arbitrarily far from equilibrium \cite{seifert2012stochastic,evans1993probability,hatano2001steady,jarzynski1997nonequilibrium,crooks1999entropy,sekimoto_langevin_1998,martinez_colloidal_2016} and has also been experimentally verified in different contexts \cite{Ciliberto2017-fu}. 
FT has several consequences, for example, it ensures the satisfaction of the second law of thermodynamics for non-equilibrium systems. Other results that follow from FT are Green-Kubo relation \cite{searles2000fluctuation}, Jarzynski equality  \cite{jarzynski1997nonequilibrium} and thermodynamic uncertainty relation \cite{hasegawa2019fluctuation}. 

Considering its natural connection, stochastic thermodynamics can get immensely benefited from the well-established field of the stochastic differential equations.  In this context, we would like to mention that in \cite{Roldan2022-hp,Chetrite2011-kp,Pigolotti2017-cx,Neri2019-yw}, the theory of martingales, which is an important concept in the theory of stochastic processes, have been exploited in understanding stochastic thermodynamics. A central result in the field of stochastic differential equations is Girsanov theorem (GT) \cite{girsanov1960transforming,oksendal2013stochastic}, which states that, for a drift change in an Ito process, the law of the new
process would be absolutely continuous with respect to the law of the original process and one can explicitly compute the associated Radon-Nikodym derivative. Many important theoretical results and applications, from particle filtering \cite{Sarkar2014-xq,Revuz1999-da} to finance \cite{Kallianpur2000-ay}, stand on GT. In this work, we employ this well-established concept of GT in the understanding of stochastic thermodynamics. Particularly, we establish a connection between FT and GT. We show that for a given Langevin equation (LE) the GT directly recovers FT as a special case. The derivation allows for the diffusion coefficient to be a function of the state and time and not necessarily a constant. We first write the GT for a Langevin equation and its time reversed counterpart. Accordingly we relate the associated forward and time reversed probability distributions via a Radon-Nikodym derivative, which is explicitly characterized using the GT. 
We then follow the formalism of Sekimoto \cite{sekimoto1998langevin} in order to associate with thermodynamic quantities and arrive at the FT for an underdamped and overdamped langevin dynamics. One may note that the derivation is valid both in transient and steady-state irrespective of the dimensionality of the system. We would also like to mention that, similarly, one can also arrive at FTs for work and entropy by writing Langevin equations in terms of the corresponding quantities. \\

The paper is structured as follows: In Sec. II, we begin by giving a brief introduction to Girsanov Theorem; in Sec. III, we discuss in detail the FT from a change of drift using the Girsanov Theorem. Finally, in Sec IV, we present a summary of the results and draw conclusions.

\section{Revisting Girsanov theorem}
 Let $X_t=X(t)\in R^n$ be a n-dimensional stochastic process, which satisfies the following stochastic differential equation (SDE), defined under a probability measure $P$. 
 \begin{equation} \label{eq01}
 dX_t=b(X_t,t)dt+\sigma(X_t,t)dB_t
 \end{equation}
Here $B_t$ is a n-dimensional Brownian motion, $b(X_t,t)$ and $\sigma(X_t,t)$ denote the drift and diffusion coefficient respectively which may be explicit functions of $X_t$ and $t$. Say, under a different measure $Q$ the drift changes to $\tilde{b}(\tilde{X}_t,t)$. 
  \begin{equation} \label{eq02}    
  d\tilde{X}_t=\tilde{b}(\tilde{X}_t,t)dt+\sigma(\tilde{X}_t,t)d\tilde{B}_t
 \end{equation}
 The Girsanov theorem states that, the Radon-Nikodyn derivative is given by, 
\begin{eqnarray}\label{eq03}
		 dQ/dP=\exp{\left(-\xi_t-\frac{1}{2}\left[\xi_t,\xi_t\right]\right)}=M_t
	\end{eqnarray}
 where, 
 \begin{equation}
   \xi_t=\int_0^t (b-\tilde{b})\sigma^{-1} dB_\tau.
 \end{equation} 
 Here $M_t$ is an exponential martingale. The term $\left[\xi_t,\xi_t\right]$ is the quadratic variation of $\xi_t$, which is assumed to satisfy Novikov's condition, i.e., $E_P(\left[\xi_t,\xi_t\right])<\infty$. More details may be found in \cite{girsanov1960transforming,oksendal2013stochastic}. In the next section, we show that, if the Langevin equation under a new measure $Q$ corresponds to the time-reversed trajectory of the original trajectory, we arrive at the FT. 
\section{Fluctuation theorem from a change of drift}
Consider the stochastic dynamics of a colloidal particle of mass $m$ in a thermal bath of strength $\sigma^2=2 \gamma T$, $T$ is the temperature and $\gamma$ is the viscosity coefficient in one dimension. We set Boltzmann`s constant $k_B$ to unity to make entropy dimensionless. Here we consider the underdamped case where the inertial effects are not suppressed by damping. The position and velocity of the colloidal particle at time $t$ are denoted by $x(t)$ and $v(t)$ respectively. The particle experiences a conservative potential $V\left(x\right)$ and a time-dependent external force $f\left(t\right)$. We represent the phase space probability density of the particle at time $t$ by $\rho(x(t),v(t);t)$.
The probability density that a colloidal particle assumes a position $x\left(t\right)$ at time $t$ given it was initially at  $x\left(0\right)$ and with velocity $v\left(0\right)$ is  $\rho_F:=\rho_F(x(t),v(t),t|x(0),v(0),0)$. Over initial time $t=0$ and final time $t=t_f$, the forward Langevin dynamics of the colloidal particle may be written as, 
	\begin{equation} \label{eq04}
		\begin{aligned}
			& d x=v dt \\
			&dv=\frac{1}{m}\left(-\gamma v-\partial_{x}V\left(x\right)+ f\left(t\right)\right)dt+\frac{\sigma}{m} d B_t
		\end{aligned}
	\end{equation}
	where $B_t$ is a Wiener process that characterizes the noise term. With respect to equation \eqref{eq01}, the above equation corresponds to $X_t=\{x, v\}$, $b(X_t,t)=\{v,\frac{1}{m}\left(-\gamma v-\partial_{x}V\left(x\right)+ f\left(t\right)\right)\}$ and $\sigma(X_t,t)=\{0, \frac{\sigma}{m}\}$. Accordingly, in the LE for the time-reversed trajectory defined as $\tilde{X}_t=\{\tilde{x}, \tilde{v}\}$, the drift term can be written as $\tilde{b}(\tilde{X}_t,t)=\{ \tilde{v}, \frac{1}{m}(\gamma \tilde{v}-\partial_{\tilde{x}}V\left(\tilde{x}\right)+ \tilde{f}\left(t\right)) \}$, and $\sigma(\tilde{X}_t,t)=\{0, \frac{\sigma}{m}\}$, where $\tilde{x}\left(t\right):=x\left(t_f-t\right)$, $\tilde{v}\left(t\right):=-v\left(t_f-t\right)$ and $\tilde{f}\left(t\right):=f\left(t_f-t\right)$ represent the position, velocity and external deterministic forcing for the time reversed trajectory respectively. $\tilde{B}_t$ is a Wiener process characterizing the noise for the reverse trajectory. The forward and the time-reversed trajectories must obey the following conditions: $x\left(0\right)=\tilde{x}\left(t_f\right)$ and $x\left(t_f\right)=\tilde{x}\left(0\right)$. The particle assumes the time-reversed trajectory with probability density $\rho_R:=\rho_R(\tilde{x}(t),\tilde{v}(t),t|\tilde{x}(0),\tilde{v}(0),0)$ given the initial position of the particle is $\tilde{x}(0)$ and initial velocity is $\tilde{v}(0)$. Using eqn \eqref{eq03}, we obtain
    \begin{eqnarray}\label{eq05}
		\frac{\rho_R}{\rho_F}=M_t
	\end{eqnarray}
	\begin{eqnarray}\label{eq06}
		\rho_R=\exp{\left(-\xi_t-\frac{1}{2}\left[\xi_t,\xi_t\right]\right)} \rho_F
	\end{eqnarray}
	where 
 \begin{equation}
      \xi_t=\int_0^t\left(-\frac{2 \gamma v+\left(\tilde{f}\left(\tau\right)-f\left(\tau\right)\right)}{\sigma}\right) dB_\tau. 
 \end{equation}

 In Appendix A, we have given the detailed steps for arriving at the expression for $M_t$ given in \eqref{eq07}. Now we identify $M_t$ in terms of thermodynamic quantities and can rewrite it as,
	\begin{eqnarray}\label{eq07}
		M_t & = & \exp \left( \int_{0}^{t}\dfrac{\sigma v}{T} dB_{\tau}-\int_{0}^{t}\dfrac{\gamma v^{2}}{T}d\tau \right) \\
		& = & \exp \left( \dfrac{Q(t)}{T} \right) \nonumber\\
		& = & \exp \left( -\Delta S_{env} \right) \nonumber\\
	\end{eqnarray}
	Here $Q(t)=\int(-\gamma v)vdt+\int\sigma vdB_{t}$ represents the heat transferred by the environment to the particle in time $t$ \cite{sekimoto1998langevin}.  Hence the associated entropy is given by, $\Delta S_{env}= -\dfrac{Q(t)}{T}$. 
	The work done to the particle by the external force $f(t)$ is given by $W(t)=\int_{x(0)}^{x(t)}f(\tau) dx(\tau)$. One must note here that the stochastic work and stochastic heat are not state functions i.e., they depend on each realization in the ensemble. We obtain the first law by equating sum of the stochastic heat and stochastic work to the change in the energy of the particle at time $t$:
	\begin{equation}\label{eq08}
		Q(t)+W(t)=\triangle E(t)=E(t)-E(0)
	\end{equation}
	
	where energy at time $t$ is given by  
	\begin{equation}\label{eq09}
		E(t)=\dfrac{1}{2}mv(t)^{2}+V(x(t))
	\end{equation}
	

	Let $\rho_0$ and $\rho_t$ be the initial and final probability distributions.
	Upon multiplying both sides of Eqn.\eqref{eq06} with $\frac{\rho_t}{\rho_0}$ and using eqn \eqref{eq07}, we get,
	\begin{equation}\label{eq10}
		\rho_t\rho_R = \rho_0\rho_{F}\exp\left(-\ln\dfrac{\rho_0}{\rho_t}+\dfrac{Q_{t}}{T}\right) 
	\end{equation}
	Here $\rho_0\rho_{F}$ and $\rho_t\rho_{R}$ are the forward and backward probabilities, $P_F$ and $P_R$ respectively. Note that $\rho_F$ and $\rho_R$ are transition kernels for the forward and the time reversed paths. 

 Therefore, from eqn \eqref{eq10}, we obtain,
	\begin{equation}\label{eq11}
		P_{R}=P_{F}\exp\left(-\left(\Delta S_{sys}+\Delta S_{env}\right)\right)=P_{F}\exp\left(-\left(\Delta S\right)\right) 
	\end{equation}
 where $\Delta S_{sys}=\ln \frac{\rho_0}{\rho_t}$ is the system entropy and the total change in entropy is $\Delta S=\left(\Delta S_{sys}+\Delta S_{env}\right)$. Thus we arrive at the FT,
	\begin{equation}\label{eq12}
		\frac{P_F}{P_R}=\exp{\left(\Delta S\right)}.
	\end{equation}
	
	On taking expectation on both sides we get the integral Fluctuation Theorem as given below:
	\begin{equation}\label{eq13}
		\big \langle \exp{\left(-\Delta S\right)}\big\rangle=1.
	\end{equation}
	%


	
Next, we move on to the case of Langevin dynamics of an overdamped colloidal particle. The forward overdamped Langevin dynamics of a colloidal particle may be written as: 
	\begin{equation}\label{eq14}
		dx=\dfrac{1}{\gamma} \left( -\dfrac{\partial V\left( x\right) }{\partial x}+f\left( t\right) \right) dt+\dfrac{\sigma}{\gamma} dB_{t}
	\end{equation}

The above equation corresponds to $X_t=x$, $b(X_t,t)=\frac{1}{\gamma}(-\dfrac{\partial V(x) }{\partial x}+f(t))$ and $\sigma(X_t,t)=\frac{\sigma}{\gamma}$ with respect to equation \eqref{eq01}. Accordingly, the LE for the time-reversed trajectory corresponds to $\tilde{X}_t=\tilde{x}$, the drift term $\tilde{b}(\tilde{X}_t,t)=\frac{1}{\gamma}(\dfrac{\partial V(\tilde{x}) }{\partial \tilde{x}}-\tilde{f}(t))$, and $\sigma(\tilde{X}_t,t)= \frac{\sigma}{\gamma}$, where $\tilde{x}\left(t\right):=x\left(t_f-t\right)$, and $\tilde{f}\left(t\right):=f\left(t_f-t\right)$ represents the position and external deterministic forcing for the time-reversed trajectory. 
Therefore, for the overdamped case, we have 

\begin{equation}
    \xi_{t}=\int ^{t}_{0} \left(- \frac{2}{\sigma }\frac{\partial V(x)}{\partial x}+\frac{1}{\sigma }(f( \tau) +\tilde{f}( \tau )) \right) dB_{\tau}
\end{equation}
		
Thus,
\begin{eqnarray}\label{eq15}
		M_t & = & \exp \left( \int_{0}^{t}\dfrac{ \sigma v}{T} dB_{\tau}-\int_{0}^{t}\dfrac{\gamma v^{2}}{T}d\tau \right) \\
		& = & \exp \left( \dfrac{Q(t)}{T} \right) \nonumber\\
		& = & \exp \left( -\Delta S_{env} \right) \nonumber\\
	\end{eqnarray}
	where, as mentioned before, $Q(t)=\int(-\gamma v)vdt+\int\sigma vdB_{t}$ represents the heat transferred by the environment to the particle in time $t$. The detailed steps for arriving at the expression for $M_t$ in eqn \eqref{eq15} is given in Appendix B. Next, we can follow similar steps as shown in the under-damped case eqns \eqref{eq10}, \eqref{eq11}, \eqref{eq12}, \eqref{eq13} and recover the FT for the over-damped case.

\section{Discussion and conclusion} 
In this work, we have shown that FT can be directly arrived as a special case of Girsanov theorem. Specifically, the Girsanov transformation of measures between the forward and the reversed Langevin dynamics of a colloidal particle recovers FT. The derivation allows for the diffusion coefficient to be a function of state and time. The derivation is also applicable to both transient and steady states. We have recovered the FT for both underdamped and overdamped Langevin equations in the similar manner. Overall, the main contribution of this work is, perhaps, discovering that FT is a special case of Girsanov theorem under an appropriate transformation of measures. This provides a strong measure theoretic foundation to stochastic thermodynamics, which is likely to have far reaching consequences.


\section{Acknowledgment}

We thank \'Edgar Rold\'an for useful comments on a preliminary draft of this manuscript. AD acknowledges the financial support from DST Inspire Faculty Grant (DST/INSPIRE/04/2020/000928), Government of India. 

\appendix
\section{Derivation of $M_t$ for the under-damped case}
In the under-damped case, we have
	\begin{eqnarray}
	\label{eq16}
		-\xi_t&=&\int_0^t\frac{2\gamma v+ \left(\tilde{f}\left(\tau\right)-f\left(\tau\right)\right)}{\sigma}dB_\tau \nonumber \\
        &=&\int_0^t\frac{2 \gamma v \sigma }{2\gamma T}dB_\tau+ \int_0^t\frac{ \left(\tilde{f}\left(\tau\right)-f\left(\tau\right)\right) }{\sigma}dB_\tau \nonumber \\
	\end{eqnarray}
	The second term on the right-hand side (RHS) can be shown to be zero. To see that, let us consider a time discretization with equal time step $\Delta \tau$,  $0=\tau_0<\tau_1...<\tau_i...<\tau_{N+1}=t$, such that $\Delta \tau=\Delta \tau_i:=\tau_{i+1}-\tau_i \, \forall \, i\in\{0,1,2,...,N\}$. The associated Brownian increments are $\Delta B_{\tau_i}:=\left(B_{\tau_{i+1}}-B_{\tau_{i}}\right)$. We may write the Ito integral as,
$$\int_0^t\frac{ \tilde{f}\left(\tau\right) }{\sigma}dB_\tau=\frac{1 }{\sigma}\sum_{i=0}^N\tilde{f}\left(\tau_i\right)\left(B_{\tau_{i+1}}-B_{\tau_{i}}\right)$$ 
Similarly, 
$$\int_0^t\frac{ f\left(\tau\right) }{\sigma}dB_\tau=\frac{1 }{\sigma}\sum_{i=0}^Nf\left(\tau_i\right)\left(B_{\tau_{i+1}}-B_{\tau_{i}}\right)$$
Therefore, we have
\begin{eqnarray}\label{eq17}
&&\int_0^t\frac{ \left(\tilde{f}\left(\tau\right)-f\left(\tau\right)\right) }{\sigma}dB_\tau \nonumber \\ 
&=&\frac{1 }{\sigma}\sum_{i=0}^N\tilde{f}\left(\tau_i\right)\Delta B_{\tau_i}-\frac{1 }{\sigma}\sum_{i=0}^Nf\left(\tau_i\right)\Delta B_{\tau_i}\nonumber \\
&=&\frac{1 }{\sigma}\sum_{i=0}^N\left(\tilde{f}\left(\tau_i\right)\Delta B_{\tau_i}-f\left(\tau_{N-i}\right)\Delta B_{\tau_{N-i}}\right) \nonumber \\
&=&0 \nonumber \\
\end{eqnarray}
	since by construct, $\tilde{f}\left(\tau_i\right)=f\left(\tau_{N-i}\right)\forall i$. Thus Eqn.\eqref{eq16} can be written as, 
	\begin{equation}\label{eq18}
		-\xi_t=\int_0^t\frac{\sigma v}{T} dB_\tau.
	\end{equation}
	and quadratic variation of $\xi_t$ can be written as,
	\begin{eqnarray}\label{eq19}
		[\xi_t,\xi_t]&=&\int_0^t{\left(\frac{\sigma v}{T}\right)^2dt} =\int_0^t{\frac{2v^2 \gamma}{T}d\tau} \nonumber \\
	\end{eqnarray}

\section{Derivation of $M_t$ for the over-damped case}

In the overdamped case, we have
\begin{equation}\label{eq20}
\xi_{t}=\int ^{t}_{0} \left[- \dfrac{2}{\sigma }\dfrac{\partial V\left( x\right) }{\partial x}+\dfrac{1}{\sigma }\left( f\left( \tau \right) +\tilde{f}\left( \tau \right) \right) \right] dB_{\tau}
\end{equation}
	 
 The second term of the integrand can be written as,
  \begin{eqnarray}\label{eq21}
&&\int_0^t\frac{ \tilde{f}\left(\tau\right) }{\sigma}dB_\tau+\int_0^t\frac{ f\left(\tau\right) }{\sigma}dB_\tau \nonumber\\
&=&\frac{1 }{\sigma}\sum_{i=0}^N\tilde{f}\left(\tau_i\right)\Delta B_{\tau_i}+\frac{1 }{\sigma}\sum_{i=0}^Nf\left(\tau_i\right)\Delta B_{\tau_i}\nonumber \\
&=&\frac{1 }{\sigma}\sum_{i=0}^N\left(f\left(\tau_{N-i}\right)\Delta B_{\tau_{N-i}}+f\left(\tau_i\right)\Delta B_{\tau_i}\right) \nonumber \\
&=&\frac{2}{\sigma}\sum_{i=0}^N f\left(\tau_i\right)\Delta B_{\tau_i} \nonumber \\
\end{eqnarray}  
Thus Eqn.\eqref{eq20} can be written as,
\begin{eqnarray}\label{eq22}
		\xi_{t}&=&\dfrac{2}{\sigma } \int ^{t}_{0} \left[ -\dfrac{\partial V\left( x\right) }{\partial x}+ f\left(\tau_i\right) \right] dB_{\tau} \nonumber\\
        &=& -\int ^{t}_{0}\dfrac{\sigma v}{T} dB_{\tau} \nonumber \\	
        \end{eqnarray}
For arriving at the expression in Eqn. \eqref{eq22}, we have used the independence property of Brownian increments. 
	Quadratic variation of $\xi_t$ can be written as,
	\begin{equation}\label{eq23}
		[\xi_t,\xi_t]=\int_0^t{\left(\frac{\sigma v}{T}\right)^2dt}=\int_0^t{\frac{2\gamma v^2}{T}d\tau}
	\end{equation}


\bibliography{mybibfile}

\end{document}